# Discovery and application of Frequency Ratio Method to the new multiperiodic γ Dor star HD 218427[*,**]

E. Rodríguez[1], P.J. Amado[1,2], J.C. Suárez[1,3***], A. Moya[1,3], M.A. Dupret[3], E. Poretti[4], A. Grigahcène[5], V. Costa[1] and M.J. López-González[1]

[1] Instituto de Astrofísica de Andalucía, CSIC, P.O. Box 3004, E-18080 Granada, Spain
[2] European Southern Observatory, Alonso de Cordova 3107, Santiago 19, Chile
[3] LESIA, Observatoire de Paris-Meudon, UMR 8109, 92190 Meudon, France
[4] INAF-Osservatorio Astronomico di Brera, Via E. Bianchi 46, 23807 Merate, Italy
[5] CRAAG, Algiers Observatory, BP 63 Bouzareah 16340, Algiers, Algeria

**Abstract.** γ Dor-type oscillations have been discovered in the star HD 218427 through simultaneous *uvby* photometric observations carried out in the year 2003. A few H$_\beta$-Crawford measurements were also collected for calibration purposes which place this star well-located inside the γ Dor instability region. Deficiency in metal content, similar to other well-defined γ Dor stars, is found in HD 218427 and the possibility of a λ Boo nature is discussed. Frequency analysis was carried out for different filters, the combined "vby" filter was also used and five frequencies were found as significant with periods ranging between 0.3 and 0.8 days. The recently developed Frequency Ratio Method is used in order to perform an identification of the excited modes. The results are consistent with an $l$=2 identification for all the modes and high radial quantum numbers ($n$∼40) for the three main observed periodicities. The possibility of multiplet structures is also discussed. However, no consistency is found when the Time-Dependent Convection treatment is used for modes discrimination. The disagreement can be due to the large rotation velocity taking place in HD 218427 and, consequently, the significant coupling between the modes.

**Key words.** stars: variables: γ Dor – stars: individual: HD 218427 – stars: oscillations – techniques: photometric

## 1. Introduction

A large number of the presently known low amplitude pulsating stars in the lower part of the Cepheid Instability Strip have been discovered while used as check stars in photometric campaigns devoted to other well-known variables. Long-term monitoring of high quality photometry is continuously showing the existence of new variables in this part of the H-R diagram. This is the case for HD 218427 which was used as second comparison star during the two-site photometric campaign carried out on the triple-mode high-amplitude pulsator AC And, prototype of the variables with the same name (Rodríguez & Breger 2001). During this run, HD 218427 (V=8.$^m$17, ST=F0, Simbad 2005) revealed a slight variability with peak-to-peak am-plitude of ΔV∼0.$^m$01 and main period of ∼0.8 days. This, together with its multiperiodic behaviour and photometric information available in the bibliography, let us assume this star as a new multiperiodic γ Dor variable and, hence, a good target to study.

The γ Dor-type variables constitute a relatively recent recognized class of pulsating variables in the zone where the red edge of the δ Sct region intersects with the main sequence. In fact, both regions are partially overlapped. More than one hundred stars in this zone are presently proposed as candidates to this group, but only a few tens are considered as *bona fide* members (Kaye et al. 1999, Handler & Shobbrook 2002, Henry & Fekel 2002, 2003, Mathias et al. 2004). Observationally, their variations are characterized by relatively long periods, ranging from about 0.3 to 3 days, and small amplitudes, between a few millimagnitudes to some hundredths, which comes from excited nonradial gravity ($g$) modes of high radial order ($n$) and low spherical degree ($l$).

The discovery of new γ Dor-type pulsators is of crucial interest in order to increase the sample of *bona fide*

*Send offprint requests to:* E. Rodríguez. E-mail: <eloy@iaa.es>
 * Table 1 is only available in electronic form at CDS
 ** Based on observations collected at Sierra Nevada and San Pedro Mártir Observatories
 *** Associate researcher at (3), with financial support from Spanish Consejería de Innovación, Ciencia y Empresa from the Junta de Andalucía local government

members, performing a reliable definition of the observational borders of this region and improving our knowledge on the constraints taking place in these variables (from various points of view: pulsational behaviour, pulsational parameters, photometric parameters, anomalies, etc).

It is also very important to gain insights on the relationship to other stars located near the same zone of the H-R diagram, with special attention on the δ Sct-type variables to whom their pulsation region partially overlap. In fact, stars exhibiting both types of pulsations have been claimed in some cases. One example is HD 209295 (Handler & Shobbrook 2002), but its γ Dor nature was not confirmed by later investigations (Handler et al. 2002). The longer variations seem to be produced by binarity effects. Recently, HD 8801 has been proposed as the first clear example of such kind of pulsators with simultaneously γ Dor and δ Sct-type variations (Henry & Fekel 2005). From a theoretical point of view, the double pulsational nature in this region has also been recently predicted by the theory (Dupret et al. 2005a).

On the other hand, the recently developed Frequency Ratio method (FRM) (Moya et al. 2005a) constitutes an important step toward the modal identification for γ Dor stars. This method, based on the first order asymptotic g-mode expression given by Tassoul (1980) is particularly useful when at least three oscillation frequencies are observed. These three frequencies, in addition with the photometric error box (temperature, luminosity and metallicity), provide an identification of the radial order and degree of the observed modes, as well as an estimate of the integral of the Brunt–Väisälä frequency weighted over the stellar radius along the radiative zone. This makes possible a large restriction on the mass, overshooting, temperatures and metallicities of the models fulfilling the complete constraints. Such a restriction is one of the main goals of asteroseismology, i.e., to sound stellar interiors by means of pulsations.

In a second paper (Suárez et al. 2005), the authors discuss the applicability of the method for rotating γ Dor variables, claiming its robustness for moderately-rotating objects ($vsini \lesssim 70$ km s$^{-1}$). In addition, improvements coming from the pulsational behaviour of γ Dor stars are likely to complete this method. Particularly, constraints on unstable modes considering the convection-pulsation interaction can be obtained from recent theoretical developments using a Time-Dependent Convection (TDC) treatment (Grigahcène et al. 2005a, Dupret et al. 2005a). In the frame of the latter works, amplitude ratios and phase shifts between light curves obtained in different filters can be predicted and compared with those observed for each mode. Thus, both FR and TDC methods can be used independently to check the results or as complementary to constrain in a more secure way the possible set of solutions proved by the FRM.

Up to date, the FRM has been applied to the stars HD 12901 (Moya et al. 2005a) and HD 48501 (Suárez et al. 2005) with satisfactory results. Although most of the presently known γ Dor stars are multiperiodic with a large number of them pulsating in probably three or more independent pulsation modes, only for a few stars there are available long time series of photometric data to perform deep analyses of their pulsational content. Moreover, long observational runs in multicolour photometry (with observations in at least three filters) and precision enough to obtain reliable amplitude ratios and phase shifts (for at least the three main pulsation frequencies) are only available for very few γ Dor-type variables. In particular, in Strömgren uvby photometry, this exists only for the variables: γ Dor itself (Balona et al. 1994), 9 Aur (Zerbi et al. 1997) and, probably, HR 8799 (Zerbi et al. 1999). The TDC was successfully applied to the two former variables by Dupret et al. (2005b) together with the stars HD 12901 and HD 48501 for which seven-colour Geneva photometry was available. In addition, the complete FRM+TDC scheme has been recently applied to analyse 9 Aur by Moya et al. (2005b) resulting in an unique asteroseismology model as final solution.

In the present work, we show the results obtained for the new discovered multiperiodic γ Dor star HD 218427 using long time series of simultaneous uvby photometry. This should be a new case in which the complete FRM+TDC procedure can be applied. Simultaneity presents great advantages concerning speed and precision in the measurements collected, avoiding problems derived from interpolation or small transparency changes in the atmosphere. This produces more precise data. This precision is very important in the determination of the existing phase shifts and amplitude ratios between different filters. However, the very small amplitudes of HD 218427 constitute a handicap in deriving these phase shifts because the error bars are consequently enlarged. Nevertheless, the possibility to apply the FRM+TDC scheme to another case than 9 Aur is very attractive.

## 2. Observations

The observations were carried out using the 90 cm telescope at Sierra Nevada Observatory (SNO), Spain, between July and September, 2003 on 21 photometric nights and about 120 hours of useful data collected. In November, a few data were additionally obtained on 7 nights with the 1.5 m telescope at San Pedro Mártir Observatory (SPMO) in Baja California, Mexico. These data did not improve the spectral window in a substantial way, but they allowed us to get a better frequency resolution. Both telescopes are equipped with identical six-channel uvbyβ spectrograph photometers for simultaneous measurements in uvby or the narrow and wide H$_\beta$ channels (Nielsen 1983). However, the majority of the data were collected in the four uvby and only a few nights at SNO were also devoted to measure H$_\beta$ for purposes of calibrating the photometric indices. A technical problem arose with the u filter in the 1.5 m telescope which made useless the u data collected with that telescope.

The main object in this observing run was the triple-mode variable AC And with C1=HD 219874 as main com-parison star and C2=HD 218427 as check star. When the variability of the latter object was suspected, a second check star, C3=HD 218985, was used. The sequence was, generally, C1,C2,C3,Var with sky measurements every 2 or 3 cycles. The extinction corrections were based on nightly coefficients determined from C1. Then, magnitude differences of each object relative to C1 were calculated by means of linear interpolation. No sign of variability within about 2 mmag was found for C1 or C3 with the observations reported here. In fact, no pulsations are expected for these objects because both are too cool and placed outside the red border of the $\gamma$ Dor region in the H-R diagram (see below).

**Table 2.** $uvby\beta$ indices obtained for HD 218427 and comparison stars. The pairs below the star names are the number of points collected for each object in $uvby$ and $\beta$, respectively. The values given by Olsen (1996) are listed in the bottom part.

| Object | V | b-y | $m_1$ | $c_1$ | $\beta$ |
|--------|------|------|------|------|------|
| | (mag) | (mag) | (mag) | (mag) | (mag) |
| HD 218427 | 8.170 | 0.222 | 0.136 | 0.649 | 2.722 |
| (823, 77) | 8 | 5 | 5 | 12 | 12 |
| C1=HD 219874 | 8.584 | 0.331 | 0.149 | 0.426 | 2.645 |
| (846, 77) | | | | | |
| C3=HD 218585 | 8.308 | 0.310 | 0.155 | 0.482 | 2.663 |
| (160, 23) | 4 | 5 | 5 | 11 | 13 |
| HD 218427 | 8.166 | 0.220 | 0.132 | 0.656 | 2.710 |
| C1=HD 219874 | 8.587 | 0.332 | 0.153 | 0.425 | 2.651 |
| C3=HD 218585 | 8.308 | 0.312 | 0.156 | 0.477 | 2.670 |

**Table 3.** Reddening and derived physical parameters for HD 218427.

| Parameter | Value | Parameter | Value |
|-----------|-------|-----------|-------|
| E(b-y) | $0.^{m}002\pm0.01$ | $M_{bol}$ | $2.^{m}71\pm0.3$ |
| $(b-y)_0$ | $0.^{m}220\pm0.01$ | D.M. | $5.^{m}46\pm0.3$ |
| $m_0$ | $0.^{m}137\pm0.01$ | $\log L/L_\odot$ | $0.82\pm0.12$ |
| $c_0$ | $0.^{m}649\pm0.01$ | $T_e$ (K) | $6950\pm150$ |
| $\delta m_1$ | $0.^{m}041\pm0.01$ | $\log g$ | $4.10\pm0.1$ |
| $\delta c_1$ | $0.^{m}043\pm0.02$ | Age (Gyr) | $1.9\pm0.1$ |
| [Me/H] | $-0.35\pm0.1$ | $M/M_\odot$ | $1.40\pm0.1$ |
| $M_v$ | $2.^{m}70\pm0.3$ | $R/R_\odot$ | $1.76\pm0.3$ |

# 3. Photometry

To transform our instrumental magnitude differences into the standard $uvby\beta$ system, we have followed the procedure described in Rodríguez et al. (1997). A very good agreement is found between the standard differences derived from the two observatories among each other and with those determined from different catalogues available in the bibliography (Olsen 1983, 1996, Hauck & Mermilliod 1998). The data, as standard magnitude differences of Variable minus C1 versus Heliocentric Julian Date are presented in Table 1 and can also be requested from the authors.

Table 2 lists the absolute standard $uvby\beta$ indices derived for HD 218427 and the two comparison stars C1 and C3 following the method described in Rodríguez et al. (2003). The error bars listed in this table mean standard deviations of magnitude differences relative to C1. As seen, our results are in a very good agreement with the values found in the homogeneous catalogue of Olsen (1996). Similar results can be found in the lists of Olsen (1983) and Hauck & Mermilliod (1998). The results concerning AC And are not presented here because they will be the subject of another much more specific study.

Suitable calibrations available in the bibliography for $uvby\beta$ photometry were used to estimate the physical parameters of HD 218427. With this purpose, the procedure described in Rodríguez et al. (2001) was followed. The results are summarized in Table 3 with the star well situated into the $\gamma$ Dor region as shown in Fig. 1. In this figure, the claimed hybrid star HD 209295 has been excluded from the list of Handler (2002) since its $\gamma$ Dor pulsation seems to be tidally excited (see Handler et al. 2002). In deriving the luminosity of HD 218427, the photometric $M_v$ value which comes from the $uvby\beta$ indices was the only one considered because its parallax was not measured by the Hipparcos satellite (ESA 1997).

Concerning the metal content, a relatively high value of $\delta m_1 = 0.^{m}043$ is obtained which leads to the low [Me/H]=−0.35, using the Smalley's (1993) relation for metal abundances. This means Z=0.0089 for this star, assuming the traditionally adopted value of $Z_\odot=0.020$ for the Sun, or Z=0.0054 if the new derived solar abundances by Asplund et al. (2005) are considered with $Z_\odot=0.0122$. Slightly deficiency in metals seems to be a common characteristic among the $\gamma$ Dor-type variables as shown in Fig. 3 of Handler (1999). In contrast with the $\delta$ Sct-type variables, nearly none $\gamma$ Dor variable show $\delta m_1$ values higher than $0.^{m}00$, the great majority of them show values between $0.^{m}00$ and $0.^{m}03$ which means Z content between 20% larger and about 50% lower than the solar one, respectively.

On the other hand, interestingly, very similar low metallicities to HD 218427 can be found for other less well-defined $\gamma$ Dor-type variables: HR 8799 (=HD 218396) and SAO 32177 (=HD 239276) (Rodríguez et al. 2005). These authors find $\delta m_1 = 0.^{m}042$, [Me/H]=−0.36 for HR 8799 and $\delta m_1 = 0.^{m}046$, [Me/H]=−0.40 for SAO 32177. Thus, the three stars nearly range the full temperature region of the $\gamma$ Dor variables as shown in Fig. 1, but with very similar deficiencies in metal abundances. Hence, it suggests that this deficiency is not dependent on the temperature and there is not a correlated slope between $\delta m_1$ and temperature as claimed by some authors (Rodríguez & Breger 2001).

Another interesting possibility consists in the slope's hypothesis be valid and the metal deficiencies in these three stars might be probably connected with a $\lambda$ Boo nature as already Gray & Kaye (1999) found for HR 8799. In fact, from a photometric point of view, the three stars can be found within the $\lambda$ Boo region of both ($m_1$,b-y) and ([m1],β) diagrams (Gray 1988, Gray & Corbally 1993). In this scheme, these stars should constitute a small subgroup of metal-poor γ Dor variables similar to the λ Boo stars pulsating as δ Sct-type variables. In such case we are dealing with metal-poor Population I stars which show significant underabundances of metals, except for the elements C, N, O and S. These stars are also characterized by showing broad, but often shallow hydrogen-line wings and weak Mg II λ 4481 lines. The mechanism which originates these anomalies in their patterns of abundances is still subject of discussion. Most of the current theoryes suggest this phenomenon is produced by interaction between the stellar surface and its local environment. The most accepted one invokes the acretion of metal-depleted gas from circumstellar material on the stellar photosphere (Venn & Lambert 1990, Turcotte & Charbonneau 1993, Gray & Corbally 2002). Others include diffusion and mass-loss (Michaud & Charland 1986, Charbonneau 1993) or the influence of binarity (Andrievsky 1997, Faraggiana & Bonifacio 1999). However, in order to verify the true λ Boo nature of SAO 32177 and HD 218427, the analysis of high-resolution spectra are crucial to morphologically discriminate from other groups of metal-poors A–F type stars.

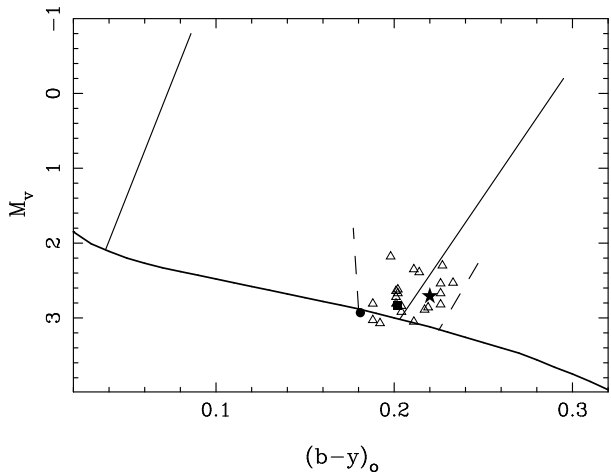

**Fig. 1.** Location of HD 218427 (star) in the H-R diagram together with the sample of *bona fide* γ Dor stars from the list of Handler (2002) and the observational edges by Handler & Shobbrook (2002). The position of the two γ Dor-type stars HR 8799 (square) and SAO 32177 (circle) are also shown having metallicities similar to HD 218427. The borders of the δ Sct region are from Rodríguez & Breger (2001).

As commonly occurs in stars inside the γ Dor-region, HD 218427 is in a main sequence stage of evolution. This is confirmed when the evolutionary tracks of Claret & Giménez (1995), with Z=0.01, are used. This way, a mass of 1.40(±0.1) $M_\odot$ and age of 1.9(±0.1) Gyr are estimated. This mass is about 12% smaller than that obtained if similar evolutionary tracks, but with Z=0.02, are considered. On the other hand, an evolutionary bolometric magnitude of $M_{bol}$=2.$^m$72 can be determined too, in very good agreement with the photometric one listed in Table 3.

**Table 4.** Frequencies, amplitudes and amplitude signal/noise ratios obtained for the combined *vby* filter.

| Frequency ($cd^{-1}$) | Amplitude (mmag) | S/N |
|---|---|---|
| | ±0.32 | |
| $f_1$=1.3326 2 | 8.13 | 14.8 |
| $f_2$=1.3591 3 | 6.81 | 12.4 |
| $f_3$=1.4779 4 | 3.58 | 6.5 |
| $f_4$=3.7164 5 | 2.38 | 4.7 |
| $f_5$=2.2813 5 | 2.30 | 4.2 |

## 4. Frequency analysis

We applied some different approaches to detect the excited frequencies. A common result was the lack of any significant peak in the high frequency domain (from 10 to 90 $cd^{-1}$). Indeed, the frequencies expected to be found in the HD 218427 light curve are lower than 4 $cd^{-1}$, assuming only independent terms corresponding to the γ Dor pulsation regime. This limit can be slightly enlarged if harmonics or combination terms play a significant role. Moreover, as seen in Fig. 1, *uvbyβ* photometry locates this star outside the δ Sct region; hence, short–period *p*–modes are not expected to be excited.

To judge the significance of a detected peak, we assumed the criterion given by Breger et al. (1993, 1999): amplitude signal to noise ratio (S/N) ≥4.0 for independent peaks. The noise level is calculated by averaging the amplitudes (oversampled by a factor of 20) over 5 $cd^{-1}$ regions around the frequency under consideration. Nevertheless, the existence of each peak is checked in the different filters as an additional test together with the relations between their amplitudes which provide us additional information on its nature. In this respect, the results obtained in the *u* filter have to be considered with caution because the data collected in this filter are the noisiest and, as mentioned in Sect. 2, the data set collected in this filter at SPMO was useless.

When analyzing the time series in each colour, we realized that the aliasing effects combined with the low-amplitude terms modified the hierarchy between the central peak and its side peaks, making the final choice not an obvious one. This is in part due to the proximity between the three main frequencies and can only be resolved with simultaneous multisite observations to minimize the power of the 1 $cd^{-1}$ aliases. In our case, this problem remains because the data at SPMO were collected two months later than at SNO.

As usual when *uvby* photometry is used to investigate the pulsational content of A–F stars, the data collected in the *vb* bands were firstly analyzed, since the amplitudes are larger than those in the *uy* bands. Three main peaks clearly appear as significant with S/N>4.0 in the $v$ data set: $f_1$=1.3328, $f_2$=1.3594 and $f_3$=1.4784 cd$^{-1}$. Similar results were found in the $b$ filter, but the aliasing effect produced some ambiguities; moreover, other terms were probably present, hidden in the noise.

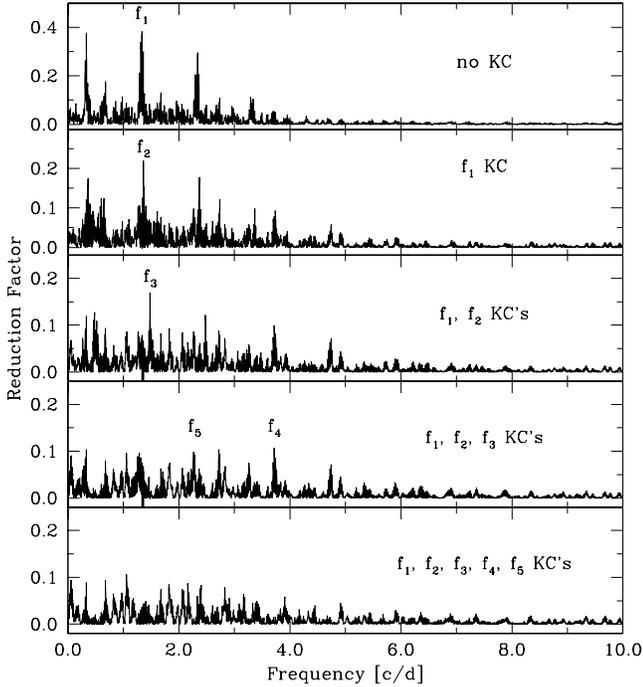

**Fig. 2.** Power spectra of HD 218427 in the combined $vby$ band corresponding to the original data and residuals after removing different sets of simultaneously optimized peaks.

**Table 6.** Observed phase shifts and amplitude ratios.

| Frequency (cd$^{-1}$) | u-y (°) | v-y (°) | b-y (°) | u/y | v/y | b/y |
|---|---|---|---|---|---|---|
| $f_1$=1.3326 | 5.2 | 2.6 | 3.2 | 1.08 | 1.53 | 1.31 |
|  | 9.7 | 5.3 | 5.5 | 15 | 11 | 10 |
| $f_2$=1.3591 | −28.2 | −4.4 | −3.1 | 0.64 | 1.32 | 1.26 |
|  | 16.6 | 6.6 | 6.5 | 15 | 11 | 11 |
| $f_3$=1.4779 | −54.1 | −20.7 | −11.0 | 0.96 | 1.14 | 0.99 |
|  | 16.9 | 10.0 | 10.2 | 26 | 17 | 15 |

In order to gain consistency in our results, a combined $vby$ filter was built by aligning the $vby$ data following the method described in Rodríguez et al. (2001). To simulate the amplitudes in $v$ band, scale factors of 1.126 and 1.376 were applied to $b$ and $y$ data taking into account the amplitudes of the three peaks previously found. Then, weights of 1.0, 1.153 and 0.732 were adopted for measurements collected in filters $v$, $b$ and $y$, respectively, in order to average the corresponding $vby$ measurements obtained at each instant.

The most straightforward results were obtained using Vaniček's (1971) least–squares technique. It searches for multiple periods without relying on prewhitening, thus avoiding dangerous deformations in the power distribution. The reduction factor $RF = 1 - \sigma_{fin}^2/\sigma_{in}^2$ is calculated for each trial frequency. $\sigma_{in}^2$ is the variance before considering the trial frequency and $\sigma_{fin}^2$ is the variance after considering it. The amplitudes and phases of the frequencies identified in the previous searches are recalculated for each trial frequency, since just the frequency values are introduced as known constituents (KC's). Thus, we always processed the original data set, without any prewhitening. The results are presented in Fig. 2 and Table 4 with 5 frequencies found as significant in the range between 1.3 to 3.7 cd$^{-1}$, typical of $\gamma$ Dor-type pulsation. None of them correspond to any combination of the other. The error bars listed in Table 4, the same for those listed in Table 5, are the formal error bars which result from the fitting; they are very similar to those which we can obtain using the analytically derived formulae by Montgomery & O'Donoghue (1999). In the top panel of Fig. 2 is shown the highest peak $f_1$=1.3328 cd$^{-1}$, the predominant term. When the $f_1$ value is introduced as KC, the $f_2$=1.3594 cd$^{-1}$ peak and its alias structure is clearly visible (second panel from top). Also the $f_3$=1.4784 cd$^{-1}$ term is quite evident (third panel from top). After that the selection between $f_4$=3.7164 cd$^{-1}$ and its alias terms is more complicated. We can suggest $f_4$=3.7164 cd$^{-1}$ as the more probable fourth term and $f_5$=2.2813 cd$^{-1}$ as a tentative fifth term. However, the scientific discussion should be restricted to the first three terms, since the last two are speculative.

These identifications are also confirmed when using the method described in Rodríguez et al. (1998) where single– and multiple–frequency techniques are combined using both Fourier and least–squares algorithms. The results were a little less obvious, since the prewhitened data slightly distorted the amplitude spectra. A further, positive check has been performed using the new available package PERIOD04 (Lenz & Breger 2005).

Table 5 lists the results when our 5-frequency solution is applied to each of the four $uvby$ filters. The results seem to be consistent, in all cases, with $\gamma$ Dor-type pulsation (Rodríguez 2005) within the present uncertainties which are particularly large in the case of the $u$ filter. Table 6 lists the corresponding phase shifts and amplitude ratios, relative to the $y$ filter, for the three main frequencies. Very different values for $f_2$ and $f_3$ relative to $f_1$ might suggest modes with different spherical quantum numbers ($l$), but the error bars are too large and this suggestion must be taken with caution. In fact, in the next section, we will assume the same $l$ for all the modes in order to apply the Frequency Ratio method. We note in particular that the phase differences observed in HD 218427 are very large for the components $f_2$ and $f_3$. This is very different from the small phase differences ($< 5°$) observed in most of the $\gamma$ Dor stars (Rodríguez 2005, Cuypers & de Cat 2005) and will be discussed below.

**Table 5.** Results from the Fourier analysis applied to the four *uvby* filters.

| Frequency (cd$^{-1}$) | $u$ A (mmag) | $\varphi$ (rad) | S/N | $v$ A (mmag) | $\varphi$ (rad) | S/N | $b$ A (mmag) | $\varphi$ (rad) | S/N | $y$ A (mmag) | $\varphi$ (rad) | S/N |
|---|---|---|---|---|---|---|---|---|---|---|---|---|
| | ±0.73 | | | ±0.35 | | | ±0.31 | | | ±0.34 | | |
| $f_1$ = 1.3326 | 5.98 | 3.171 / 113 | 6.6 | 8.48 | 3.125 / 38 | 14.1 | 7.24 | 3.136 / 40 | 14.5 | 5.53 | 3.080 / 56 | 11.5 |
| $f_2$ = 1.3591 | 3.17 | 5.744 / 225 | 3.5 | 6.57 | 6.158 / 51 | 11.0 | 6.24 | 6.181 / 48 | 12.5 | 4.97 | 6.236 / 65 | 10.4 |
| $f_3$ = 1.4779 | 2.91 | 0.948 / 235 | 3.2 | 3.46 | 1.531 / 83 | 5.8 | 3.00 | 1.701 / 86 | 6.0 | 3.04 | 1.892 / 91 | 6.3 |
| $f_4$ = 3.7164 | 2.27 | 4.089 / 242 | 3.0 | 2.23 | 4.354 / 118 | 4.2 | 2.20 | 4.335 / 108 | 4.8 | 1.79 | 4.270 / 143 | 3.7 |
| $f_5$ = 2.2813 | - | - / - | - | 2.28 | 4.078 / 118 | 3.8 | 2.01 | 4.081 / 121 | 4.1 | 1.75 | 4.257 / 149 | 3.6 |
| residuals (mmag) | 10.7 | | | 5.3 | | | 4.8 | | | 5.2 | | |

# 5. Modal identification

## 5.1. The Frequency Ratio method

The Frequency Ratio method (Moya et al. 2005a) has been applied to HD 218427 in order to obtain information on: 1) possible identification of the radial order $n$ and degree $l$ of the modes corresponding to the observed frequencies and 2) an estimate of the integral of the buoyancy frequency (Brunt–Väisälä) weighted over the stellar radius along the radiative zone ($\Im$). In order to apply this method, we assume that all the excited modes correspond to the same spherical degree $l$ with $m$=0. The additional information provided by the $\Im$ integral constitute also a reliable manner for discriminating models within the metallicity error box. Since the rotational velocity of this object ($v\sin i$=72 km s$^{-1}$) (Glebocki et al. 2000) is in the limit of validity of the FRM in presence of rotation (Suárez et al. 2005), an additional study of multiplet-like structures is performed.

### 5.1.1. Modelling

The equilibrium models are built using the evolutionary code CESAM (Morel 1997) with the physical parametrisation adapted for intermediate mass stars: the parameters $\alpha_{ML} = l_m/\mathrm{Hp} = 1.8$ and $\alpha_{ov} = l_{ov}/\mathrm{Hp} = 0.2$ are used, for the convective efficiency and the mixed core overshooting, respectively. Hp corresponds to the local pressure scale-height; $\alpha_{ov}$ represents the inertial penetration distance of convective bulbs and $l_m$ is the mixing length. The opacity tables are taken from the OPAL package (Iglesias & Rogers 1996), complemented at low temperatures ($T \leq 10^4 K$) with the tables provided by Alexander & Ferguson (1994). Finally, the atmosphere is reconstructed using the Eddington's $T(\tau)$ law (grey approximation).

A first order effect of rotation is taken into account in equilibrium models in the way described in Kippenhahn & Weigert (1990). That is, the spherical symmetric contribution of the centrifugal acceleration is included by means of an effective gravity $g_{\mathrm{eff}} = g - \mathcal{A}_c(r)$, where $\mathcal{A}_c(r) = \frac{2}{3} r\Omega^2$ represents the centrifugal acceleration of matter elements in a $r$ distance from the centre of the star. The total angular momentum of the models is assumed to be globally conserved along the evolution.

Concerning the computation of the oscillations, the adiabatic eigenfrequencies of representative models are computed with the oscillation code FILOU (see Tran Minh & Léon 1995). In this code, effects of rotation on the oscillation frequencies, up to second order, have been included (see Suárez 2002 and references therein).

### 5.1.2. Modal identification and model constrains given by the FRM

Glebocki et al. (2000) list $v\sin i$=72 km s$^{-1}$ for HD 218427. Hence, this star can be considered as a moderately-rotating object. Such rotational velocity is in the limit of applicability of the FRM as discussed in Suárez et al. (2005). Unfortunately we have not information on the true rotational velocity of this star. High-resolution spectroscopic observations should be very valuable to gain insights on the $m$ structure of the excited modes and the true rotational velocity of this star. Nevertheless, as presently we know *bona fide* $\gamma$ Dor stars with $v\sin i$ values widely spread, between close to 0 to nearly 200 km s$^{-1}$, they seem to be in general slow or moderate rotators. A mean value of 58(±39) km s$^{-1}$ can be derived by using the sample of 25 *bona fide* members of Handler (2002) with available measurements of $v\sin i$. This value decreases to 45(±18) km s$^{-1}$, if the three stars with the largest $v\sin i$ values are removed. This is much smaller than the average of 109(±58) km s$^{-1}$ found by Rodríguez et al.

(2000) for the low amplitude ($\Delta V \leq 0.^{m}03$) $\delta$ Sct-type pulsators. Hence, we feel the FRM should be applicable to HD 218427 without large uncertainties.

The FRM is here applied to the three main observed frequencies $f_1$, $f_2$ and $f_3$. Their corresponding observed frequency ratios are

$$\frac{f_1}{f_2} = 0.9805, \quad \frac{f_2}{f_3} = 0.9196, \quad \frac{f_1}{f_3} = 0.9017. \quad (1)$$

As explained in Moya et al. (2005a), we search for possible natural number ($+0.5$) ratios reproducing these observed values and fulfilling the relation of radial orders given by the asymptotic $g$ mode theory. Following these authors, only dipoles ($l=1$) and quadrupoles ($l=2$) are considered. In Table 7, all possible natural number ratios are listed. An error of $\pm 1.2 \cdot 10^{-2}$ is assumed for the calculation of possible radial order sets (details in Suárez et al. 2005). In addition, for the two most relevant mode degree values: $l=1$ and 2, an estimate of the observed Brunt–Väisälä integral $\mathcal{I}_{obs}$ is provided. This new observable allows us to obtain a first model constraint by placing it in the $\mathcal{I}_{th}$–log $T_e$ diagram described in Moya et al. (2005a), where $\mathcal{I}_{th}$ represents the theoretical Brunt–Väisälä integral calculated from equilibrium models. Considering the photometric error box of HD 218427 given in Table 3, it is shown that the observed frequencies likely belong to $l=2$ modes, since possible $\mathcal{I}_{obs}$ for $l=1$ do not reproduce the observed metallicity. In addition, this analysis predicts masses in the range of 1.30–1.40 $M_\odot$ for a metallicity of [Me/H]$=-0.35$, in good agreement with the results listed in Table 3. Taking into account the additional constraints on $\mathcal{I}_{obs}$, log $g$ and $T_e$ with the corresponding error bars, a short list of representative models (Table 8) is obtained for M=1.3 and 1.4 $M_\odot$ assuming an uncertainty of about 1% for the $\mathcal{I}$ identification. The most representative models are written in boldface in Table 8 for the two considered masses. These correspond to the following possible mode identification ($l=2$, $m=0$) for models with M=1.30 $M_\odot$:

$$f_n = \begin{cases} (43,\ 42,\ 39) & \text{for } \mathcal{I}_{th} \approx 856.1 \\ (44,\ 43,\ 40) & \text{for } \mathcal{I}_{th} \approx 877.0 \\ (45,\ 44,\ 40) & \text{for } \mathcal{I}_{th} \approx 887.4 \end{cases} \quad (2)$$

and for models with M=1.40 $M_\odot$:

$$f_n = \begin{cases} (43,\ 42,\ 39) & \text{for } \mathcal{I}_{th} \approx 852.5 \text{ or } \mathcal{I}_{th} \approx 862.4 \\ (44,\ 43,\ 40) & \text{for } \mathcal{I}_{th} \approx 873.0 \end{cases} \quad (3)$$

with $f_n = (n_1, n_2, n_3)$. These results constitute an important restriction for the modal identification and thereby for the modelling. Moreover, further constraints can be furnished if the whole set of observed frequencies is considered. As an illustration, if the observed frequencies $f_4$ and $f_5$ are included in this study, the FRM discards the second solution in both cases (Eq. 2 and 3). Hence, the list of solutions would be reduced to

$$f_n = \begin{cases} (43,\ 42,\ 39,\ 15,\ 25) & \text{for } M = 1.30,\ 1.40\ M_\odot \\ (45,\ 44,\ 40,\ 16,\ 26) & \text{for } M = 1.30\ M_\odot \end{cases} \quad (4)$$

**Table 7.** List of possible $(n, \mathcal{I}_{obs})$ identifications provided by the FRM applied to the observed frequencies of HD 218427. The first three columns represent the resulting $n$ identifications. Last two, correspond to the *observed* Brunt–Väisälä integral $\mathcal{I}_{obs}$ when assuming $l=1$ and 2, respectively.

| $n_1$ | $n_2$ | $n_3$ | $\mathcal{I}_{obs}(\ell=1)$ | $\mathcal{I}_{obs}(\ell=2)$ |
|---|---|---|---|---|
| 36 | 35 | 32 | 1240.51 | 716.21 |
| 37 | 36 | 33 | 1275.46 | 736.38 |
| 38 | 37 | 34 | 1310.40 | 756.56 |
| 39 | 38 | 35 | 1345.34 | 776.73 |
| 40 | 39 | 36 | 1380.29 | 796.91 |
| 41 | 40 | 37 | 1415.23 | 817.08 |
| 42 | 41 | 38 | 1450.18 | 837.26 |
| 43 | 42 | 39 | 1485.12 | 857.43 |
| 44 | 43 | 40 | 1520.06 | 877.61 |
| 45 | 44 | 40 | 1555.01 | 897.78 |
| 46 | 45 | 41 | 1589.95 | 917.96 |
| 47 | 46 | 42 | 1624.90 | 938.13 |
| 48 | 47 | 43 | 1659.84 | 958.31 |
| 49 | 48 | 44 | 1694.78 | 978.48 |
| 50 | 49 | 45 | 1729.73 | 998.66 |
| 51 | 50 | 46 | 1764.67 | 1018.83 |
| 52 | 51 | 47 | 1799.62 | 1039.01 |
| 53 | 52 | 48 | 1834.56 | 1059.18 |
| 54 | 53 | 48 | 1869.50 | 1079.36 |
| 55 | 54 | 49 | 1904.45 | 1099.53 |
| 55 | 54 | 49 | 1904.45 | 1099.53 |
| 56 | 55 | 50 | 1939.39 | 1119.71 |
| 56 | 55 | 51 | 1939.39 | 1119.71 |
| 57 | 56 | 51 | 1974.34 | 1139.88 |
| 57 | 56 | 52 | 1974.34 | 1139.88 |
| 58 | 57 | 52 | 2009.28 | 1160.06 |
| 59 | 58 | 53 | 2044.22 | 1180.23 |
| 60 | 59 | 54 | 2079.17 | 1200.41 |

with $f_n = (n_1, n_2, n_3, n_4, n_5)$. However, this later result must be interpreted with caution since the larger observational uncertainties of $f_4$ and $f_5$ may significantly increase the intrinsic error of the FRM.

Finally, proceeding as in Suárez et al. (2005), the possibility of the observed frequencies belonging to a rotational multiplet is examined. To do so, representative *pseudo-rotating* seismic models (with physical characteristics in the range of those listed in Table 8) are computed as described in the previous section. If the minimum observed surface velocity ($i = 90°$) is considered, i.e. 72 km s$^{-1}$, the observed average spacing (1$\mu$Hz approximately) is not reproduced by the theoretical oscillation spectra for any possible combination of the observed frequencies in a $l=1$ triplet-like structure. In contrast, the possibility of any combination of $l=1$ and 2 or m$\neq$0 cannot be discarded. However, such scenario is rather improbable (see Suárez et al. 2005).

## 5.2. Time-Dependent Convection analysis

Starting from the structure models selected by the FRM (Table 8) we have made a stability analysis of the os-cillation modes. We use for this the Time-Dependent Convection (TDC) presented in Grigahcène et al. (2005a) concerning multicolour photometry and predicted amplitude ratios versus phase shifts diagrams for different pairs of Strömgren filters, as applied to HD 218427. However, no definite conclusions have been reached when our observations (Table 6) are tested with the models. Our results show that $f_1$ and $f_2$ are consistent with pulsation modes with spherical degree $l=2$ and 1, respectively, in the (v,y) and (b,y) diagrams. Identification of $l=2$ is also true for $f_1$ and pairs including the $u$ filter, but none is consistent for $f_2$. The major point of discord is concerning $f_3$, in which the observed error boxes in the amplitude ratios versus phase shifts diagrams are clearly in disagreement with the predictions for any pair of filters. This shows that, although these recent theoretical developments drive the predictions in the right direction, some improvements are still needed to be implemented in the models to produce suitable predictions to reproduce the observations of $\gamma$ Dor-type pulsators.

**Table 8.** Characteristics of the selected 1.30 and 1.40 $M_\odot$ models representative of HD 218427 within the photometric error box. From left to right, columns represent: stellar mass (in solar units), the theoretical Brunt–Väisälä integral, effective temperature, surface gravity, stellar radius (in solar units) and mean density (g cm$^{-3}$). Those in boldface correspond to the most representative models with $\mathfrak{I}_{th}$ integrals predicted by the FRM (see Table 7).

| $M/M_\odot$ | $\mathfrak{I}_{th}$ | $\log T_e$ | $\log g$ | $R/R_\odot$ | $\bar\rho$ |
|---|---|---|---|---|---|
| 1.30 | 852.374 | 3.844 | 4.183 | 1.527 | 8.993 |
| 1.30 | **856.106** | 3.843 | 4.174 | 1.543 | 8.730 |
| 1.30 | 860.191 | 3.842 | 4.165 | 1.559 | 8.466 |
| 1.30 | 864.152 | 3.841 | 4.156 | 1.575 | 8.202 |
| 1.30 | 868.048 | 3.840 | 4.147 | 1.592 | 7.940 |
| 1.30 | 872.511 | 3.838 | 4.137 | 1.610 | 7.679 |
| 1.30 | **876.980** | 3.837 | 4.127 | 1.629 | 7.419 |
| 1.30 | 882.121 | 3.835 | 4.117 | 1.648 | 7.162 |
| 1.30 | **887.364** | 3.834 | 4.107 | 1.668 | 6.908 |
| 1.40 | 852.501 | 3.841 | 4.023 | 1.906 | 4.981 |
| 1.40 | **862.403** | 3.838 | 4.005 | 1.945 | 4.687 |
| 1.40 | **873.019** | 3.834 | 3.987 | 1.986 | 4.407 |

Other independent constraints on the models can be obtained by using the TDC non-adiabatic treatment presented in Grigahcène et al. (2005a). On the one hand, this analysis enables us to determine which modes are predicted to be unstable and the comparison with the observed frequencies gives constraints on the treatment of convection. On the other hand, the photometric amplitude ratios and phase differences between different passbands can be determined and the comparison with observations allows the identification of the degree $l$ of the modes and gives additional constraints on the models. We present here a summary of the results obtained for the structure models selected by the FRM (Table 8) as well as for other models.

We begin with the stability analysis. Contrary to what occurs in $\beta$ Cephei and SPB stars, the size of the convective envelope in $\gamma$ Dor-type variables depends much more on the mixing-length parameter than on metallicity. For all the models with mixing-length parameter $\alpha > 1.5$, the three observed modes are predicted to be unstable for $1 \leq l \leq 4$. For the models with $\alpha \leq 1.5$, some or all of the observed modes are predicted to be stable. Therefore, from the point of view of the stability analysis, $\alpha$ must be greater than 1.5. The results of our stability analysis are essentially the same for different metallicities, as was proved in a general way by Grigahcène et al. (2005b).

We consider now the photometric amplitude ratios and phase differences between different Strömgren passbands. As an example, we give in Fig. 3 the theoretical and observed amplitude ratios obtained for one of our models with $\alpha = 1.5$. We begin by considering the component $f_1$. For all the models considered, the theoretical predictions for the phase differences fall inside the observed error bars for this mode, whatever the degree $l$. More information about $l$ can be obtained from the amplitude ratios. We see in Fig. 3 that a good agreement between theory and observation can be obtained and this mode is identified most probably as an $l = 2$ mode, in agreement with the predictions of the FRM. We consider now the components $f_2$ and $f_3$. As mentioned previously, the observed phase differences for these modes are very high compared to what is generally observed in $\gamma$ Dor stars and the associated error bars are large. None of our models are able to predict such phase differences in all passbands at the same time. Concerning the amplitude ratios for $f_2$, the very low amplitude observed in the $u$ passband (also unusual compared to typical $\gamma$ Dor stars) cannot be explained by our theoretical models. If we consider only the $v$, $b$ and $y$ passbands, the best agreement with observation for $f_2$ is found for an $l = 1$ mode, but $l = 2$ cannot be eliminated. No mode identification can be achieved for $f_3$, because of the large error bars and the discrepancy between theory and observations for this mode.

The main origin for these discrepancies with TDC predictions probably comes from the unusually high phase shifts which occur in this star. The reason of such large phase shifts is still unknown, we can only do some guesses. Nevertheless, we also note that the observational error bars in these phase shifts are large. A possible explanation could come from rotation, which is not small for this star. In fact, recent works by Dupret et al. (2005b) show that TDC models succeed in reproducing the observed amplitude ratios and phase differences for some relatively slow rotating $\gamma$ Dor stars. The disagreement found in HD 218427 for the components $f_2$ and $f_3$ could be due to its large rotation velocity and the significant coupling between the modes due to rotation. Here the influence could be much larger than in the FRM approach. Studies by Daszyńska-Daszkiewicz et al. (2002) for $\beta$ Cephei stars and Townsend (2003) for SPB stars show that the coupling of the modes due to rotation can significantly affect the photometric amplitude ratios and phase differences. In particular, we emphasize the very high phase shifts predicted in some of the models of Daszyńska-Daszkiewicz et al. (2002). Up to date, no such study has been performed for $\gamma$ Dor-type variables; the rotation coupling is not included in our nonadiabatic code yet. The case of HD 218427 suggests that the treatment of the rotation in $\gamma$ Dor stars deserve more attention in future. Maybe other peculiarities in this star could be responsible of the shown discrepancy, but more observations would be required to address this point.

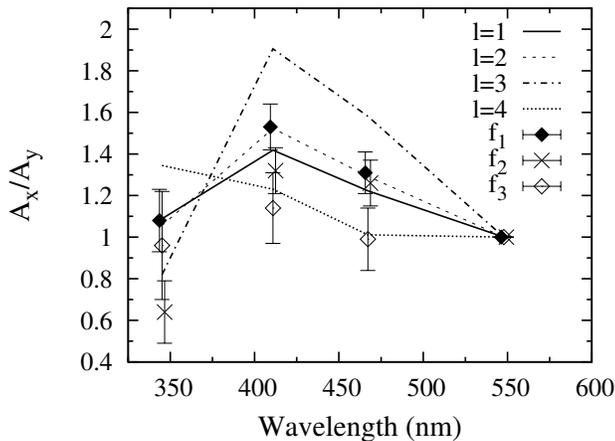

**Fig. 3.** Strömgren photometric amplitude ratios for a model of HD 218427 with $\log T_e = 3.846$, $\log g = 4.097$ and $\alpha = 1.5$. The lines are the theoretical predictions of our TDC models for different $l$. The error bars represent the observations for the frequencies $f_1$, $f_2$ and $f_3$.

*Acknowledgements.* The authors thank G. Handler for useful comments on this paper. We also thank the referee for careful examination of the manuscript and valuable suggestions which helped to improve this paper. This research was partially supported by the Junta de Andalucía and the Dirección General de Investigación (DGI) under projects AYA2003-4651 and ESP2004-03855-C03-01. JCS acknowledges the financial support of the European Marie Curie action MERG-CT-2004-513610, and the Spanish Consejería de Innovación, Ciencia y Empresa from the Junta de Andalucía local goverment. This research has also made use of the Simbad database, operated at CDS, Strasbourg, France.